\documentclass[sigconf]{acmart}
\AtBeginDocument{%
  }

\acmConference[DESTION '25]{7th International Workshop On Design Automation For CPS and IoT}{May 6,
  2025}{Irvine, CA}

\usepackage{algorithmic}
\usepackage{amsmath}
\usepackage{amsfonts}
\usepackage{algorithm}
\usepackage{array}
\usepackage[caption=false,font=normalsize,labelfont=sf,textfont=sf]{subfig}
\usepackage{textcomp}
\usepackage{stfloats}
\usepackage{url}
\usepackage{verbatim}
\usepackage{graphicx}
\usepackage{xcolor}
\definecolor{webred}{rgb}{0.5,0,0}
\definecolor{webblue}{rgb}{0,0,0.8}

\usepackage[inkscapelatex=false]{svg}
\usepackage{multirow}
\graphicspath{{Images/}}

\newtheorem{assumption}{Assumption}
\newcommand{\simulate}{S}
\newcommand{\region}{\text{region}}
\newcommand{\regionset}{\boldsymbol{\mathcal{R}}}
\newcommand{\dims}{n}
\newcommand{\dimsiter}{d}
\newcommand{\ms}{y}
\newcommand{\allms}{Y}
\newcommand{\ip}{x}
\newcommand{\pts}{\phi}
\newcommand{\ipival}{\mathcal{I}}
\newcommand{\simiter}{i}
\newcommand{\ch}{\texttt{CH}}
\newcommand{\nsims}{\kappa}
\newcommand{\randip}{g}
\newcommand{\genip}{\texttt{NextPt}}

\newcommand{\st}{\text{s.t.}}
\DeclareMathOperator*{\argmax}{\arg\!\max}

\newcommand{\dist}{\texttt{Dist}}
\newcommand{\hulldist}{\texttt{HullDist}}
\newcommand{\hypcube}{C}
\newcommand{\cubecenter}{p}
\newcommand{\cubept}{q}
\newcommand{\cuberad}{r}
\newcommand{\unitvec}{\vec{\mathbf{1}}^{(\dims)}}
\newcommand{\egregion}{R}
\newcommand{\nearestpt}{p}
\newcommand{\ptdist}{\texttt{PointDist}}
\newcommand{\optbudget}{\beta}
\newcommand{\linprog}{\text{LP}}
\newcommand{\objfunc}{f}
\newcommand{\gearnum}{\alpha}
\newcommand{\transpose}{T}
\newcommand{\distmat}{D}

\newcommand{\atextbitvector}{\textit{Extended Bit Vector}}
\newcommand{\RDM}{\text{RDM}}
\newcommand{\CRS}{\text{CRS}}
\newcommand{\exampleset}{P}
\newcommand{\cvxasm}{\text{Convex Mode Sequence Assumption}}

\begin{document}

\title[Finding Unknown Unknowns using Cyber-Physical System Simulators (Extended Report)]{Finding Unknown Unknowns using Cyber-Physical\\ System Simulators (Extended Report)}

\author{Semaan Douglas Wehbe}
\affiliation{%
  \institution{Stony Brook University}
  \city{Stony Brook, New York}
  \country{USA}}
\email{swehbe@cs.stonybrook.edu}

\author{Stanley Bak}
\affiliation{%
  \institution{Stony Brook University}
  \city{Stony Brook, New York}
  \country{USA}}
\email{stanley.bak@stonybrook.edu}

\begin{abstract}

Simulation-based approaches are among the most practical means to search for safety violations, bugs, 
and other unexpected events in cyber-physical systems (CPS).
Where existing approaches search for simulations violating a formal specification or maximizing a notion of coverage, in this work
we propose a new goal for testing: to discover unknown rare behaviors by examining discrete mode sequences.
We assume a CPS simulator outputs mode information, and strive to explore the sequences of modes produced by varying the initial state or time-varying uncertainties.
%
We hypothesize that rare mode sequences are often the most interesting to a designer,
and we develop two accelerated sampling algorithms that speed up the process of finding such sequences.
%
%
We evaluate our approach on several benchmarks, ranging from synthetic examples to Simulink diagrams of a CPS, demonstrating in some cases 
a speedup of over 100x compared with a random sampling strategy.
\end{abstract}



\maketitle

\section{Introduction}
The systems engineering process~\cite{blanchard1990systems} begins with a concept of operations, followed by requirements, design, implementation, integration and test, and fielding.
During the course of the process, detecting misbehavior or errors earlier can result in substantial savings in time and cost.
%
%
In this work we propose an additional use for integrated and CPS component simulators: finding previously unknown behaviors.

Specification-guided testing searches for \emph{known} potential problems by looking for an unknown test input (known unknowns).
In contrast, our method searches for \emph{unknown} issues with unknown test inputs (unknown unknowns).
Although not all unprecedented behaviors are necessarily problematic, we believe this process can often identify a small set of interesting candidate simulations that is feasible for manual review.
In order for this idea to work, we must have a way to detect that a simulation exhibits a rare behavior.

The goal of this work is to analyze a cyber-physical system
for which we have access to a simulator
$\simulate$.
%
%
Simulators for CPS are often designed by engineers using
languages like MATLAB Simulink~\cite{jin2014powertrain},
or via large and mostly opaque domain-specific simulators
like CARLA~\cite{dosovitskiy2017carla}.
Our accelerated testing approach does not require any visibility
into the simulator's implementation,
only its simulation outputs.
We require $\simulate$ to be a deterministic function of
$\dims$ input variables,
which may represent disturbances,
control inputs,
or initial state uncertainty.
%
%
We refer to a valuation $\ip \in \mathbb{R}^{\dims}$
of the input variables
as an \textit{input point}.
For each system, we analyze a finite interval of input points
\(
\ipival
= \left[ l, u \right]
= \left\{
    \ip \in \mathbb{R}^{n}
    \mid
    l^{(\dimsiter)} \leq \ip^{(\dimsiter)} \leq u^{(\dimsiter)}, \,
    \dimsiter=1,\,\dots,\,\dims
\right\}
\).

Tools like MATLAB Stateflow can output
information about the discrete state
of the simulation trace over time.
Given the system's set of discrete states $\Sigma$,
we define a simulation's \emph{mode sequence} to be
the sequence of discrete states visited during the simulation.
We assume that the mode sequence $\ms$ is the output
of the simulator,
$\simulate(\ip) = \ms \in \Sigma^{*}$.
%
%
We assume that
each mode sequence represents a distinct behavior of the system,
and that any two mode sequences can be compared for equality.

Let
$\allms = \{ \ms \in \Sigma^{*} \mid \exists \ip \in \ipival, \, \simulate(\ip) = \ms \}$
be the set of all mode sequences
that can be produced by simulating the CPS of interest.
For all $\ms \in \allms$,
we construct a mapping from the mode sequence $\ms$ to its preimage,
the set of input points that produce $\ms$ upon simulation:
$\pts(\ms) = \{ \ip \in \ipival \mid \simulate(\ip) = \ms \}$.
The input space $\ipival$ can be partitioned by mode sequence,
\(
    \ipival = \bigcup_{\ms \in \allms} \pts(\ms)
\).
%
Random simulations may waste much of the simulation budget
by running simulations that repeat common
or expected behaviors.
We propose an accelerated testing method that finds rare mode sequences
in fewer simulations than random sampling,
allowing
potentially unexpected or incorrect behaviors
to be discovered earlier in the process.
This work makes the following assumption about mode sequences and their
corresponding input points:
\begin{assumption}[$\cvxasm$]
\label{assumption_convex}
\[
    \forall \ms \in \allms,
    \enspace
    \pts(\ms) = \ch(\pts(\ms))
\]
\end{assumption}
where $\ch$ is the convex hull operator
\begin{equation}
\ch(\exampleset) = \{ \lambda p + (1 - \lambda) q \mid p, q \in \exampleset, \, \lambda \in [0, 1] \}
\end{equation}
Intuitively,
Assumption~\ref{assumption_convex}
makes sense
in that nearby input points should behave similarly,
up to some boundary where simulations switch from
producing one mode sequence to another.
%
%
Although Assumption \ref{assumption_convex} is
unlikely to be strictly true in real CPS with many possible behaviors
and large input spaces,
in practice,
it offers a useful criterion by which to
select input points that
efficiently discover rare mode sequences.
Of the first $\simiter$ input
points, we refer to the subset that produces mode sequence
$\ms$ as
\(
    \pts_{\simiter}(\ms) =
    \{
        \ip_j
        \mid \,
        \simulate(\ip_j) = \ms,\,
        j \in [1, \simiter]
    \}
\)
We define $\allms_{\simiter}$ to be the set of
all mode sequences
produced by at least one of the first $\simiter$ simulations.
Assumption~\ref{assumption_convex} gives
us the following property:
\begin{equation}
    \label{eq_in_ch_known_ms}
    \forall \ip \in \ch(\pts_i(\ms)),
    \enspace
    \simulate(\ip) = \ms
\end{equation}
Therefore, there is no need to simulate any
input point
$\ip$ that lies inside the convex hull
of an existing set of simulated input points
$\pts_i(\ms)$.
We refer to $\egregion_\ms = \ch(\pts_{\simiter}(\ms))$ as a
\textit{mode sequence region},
or simply the \textit{$\region$},
corresponding to the mode sequence $\ms$
after $\simiter$ simulations.
Each $\region$ constitutes a convex subset
of the input space
whose elements behave the same as one another.
%
%
We define
\(
\regionset_{\simiter} =
\{
\ch(\pts_{\simiter}(\ms_j))
\mid
j \in \left[ 1, |\allms_{\simiter}| \right]
\}
\)
to be the set of all $\region$s
after $\simiter$ simulations.


Given a finite budget of $\nsims$ simulations,
our goal is to
maximize the number of
distinct mode sequences
$|\allms_{\nsims}|$
produced by
$\simulate(\ip_1), \, \dots, \, \simulate(\ip_{\nsims})$.
%
Our accelerated testing approach takes advantage of 
Assumption \ref{assumption_convex}
to bypass simulations with familiar mode sequences.
By leveraging our knowledge of prior simulation outcomes,
we can
focus our simulation effort on promising input points,
allowing us to discover
rare and potentially unexpected behaviors
earlier in the testing process.

\section{Related Work}

Many techniques have been proposed to analyze CPS models, ranging from verification to testing approaches.
Formal methods such as reachability analysis~\cite{althoff2021set,chen2022reachability} compute sets of states that a hybrid system can enter in bounded time, providing strong guarantees about system behaviors.
These methods usually require white-box information about symbolic differential equations, and are not applicable for most large CPS simulators that are given either in Simulink or as custom simulators in large code bases.
Discrepancy function approaches~\cite{fan2015bounded,duggirala2015c2e2} can be used to provide probabilistic guarantees for systems by sampling trajectories from a black-box simulator.
This process has been extended to gray-box simulators~\cite{Fan2017Dryvr:Systems} where the discrete mode behavior of a simulator is known, as well as to control synthesis problems~\cite{qi2018dryvr}.
The assumptions for these approaches are stronger than our work, since they require knowledge of the mode transition graph and symbolic switching conditions, whereas we only require mode information as an output signal over time.
%

When specifications are given in a formal language like Signal Temporal Logic (STL)~\cite{donze2010robust}, falsification techniques can use general optimization algorithms to search for counterexamples.
These methods convert a system trace to a scalar robustness score~\cite{fainekos2009robustness}, seeking traces that minimize the robustness and violate the specification.
Like our work, tools such as S-TaLiRo~\cite{annpureddy2011s} and Breach~\cite{donze2010breach} search over initial state and time-varying uncertainty.
Falsification approaches can be effective, but they require the user to provide an STL specification, unlike our approach.
%

Several approaches use partial simulation segments to explore CPS behaviors.
One line of work builds on the Rapidly-Exploring Random Tree algorithm~\cite{lavalle2006planning} from robot motion planning in order to search for safety violations~\cite{plaku2009hybrid,kim2005rrt,dreossi2015efficient}.
Multi-shooting methods~\cite{zutshi2014multiple}, in contrast, use partial simulations to construct a discrete abstraction of a system and bridge gaps between partial simulations using abstraction refinement.
To construct simulation segments, these approaches require simulators that are fully observable and can be started and stopped from arbitrary states, which may be inapplicable for many large domain-specific simulators such as CARLA.

Software testing schemes often strive to search for inputs that improve code coverage, with common metrics being line coverage, condition coverage, or Modified-Condition Decision Coverage (MCDC).
For CPS control software, commercial tools like Reactis~\cite{ReactisProductDescription} can automatically run tests to improve MCDC, although these only analyze software.
Another approach~\cite{sheikhi2022coverage} adapts fuzz testing methods from software to CPS, by defining a notion of coverage related to the physical variables.
This method is also specification-free, but unlike our work, it ignores discrete mode information from the simulator and focuses only on coverage in a continuous space.

Rare event simulation methods~\cite{bucklew2004introduction} strive to accurately estimate the probability of rare events using the importance sampling technique from statistics. 
These methods have been used for high fidelity autonomous driving simulators~\cite{o2018scalable}, but they require a continuous measure of safety, similar to a robustness score in STL, to direct the search for known rare events.
\section{Accelerated Testing Algorithm}

In this section,
we describe two strategies for accelerated testing: Convex Rejection Sampling (CRS) and Region Distance Maximization (RDM).
Each method uses
the assumption that $\region$s are convex sets of input points
with identical behavior.
By leveraging this assumption,
these methods avoid redundant simulations,
focusing simulator effort on parts of the input space where 
it is possible to increase
the number of distinct mode sequences
$|\allms_{\nsims}|$.


\subsection{Convex Rejection Sampling}

The first proposed procedure, called \emph{Convex Rejection Sampling}, selects new input points
that lie outside all existing $\region$s using rejection sampling.
$\CRS$ begins by choosing a candidate input point
$\ip_\simiter \in \ipival$ at random.
We then perform $|\regionset_{\simiter-1}|$
convex hull containment checks
to see if $\ip_\simiter$ lies within any of the existing
$\region$s.
Each containment check is formulated
as a linear program ($\linprog$).
Because $\linprog$s scale efficiently to high dimensions,
our accelerated testing approach can be applied to
CPS simulators with high-dimensional input spaces.
Any candidate $\ip_\simiter$ that lies within a $\region$
is rejected,
since we know by Assumption \ref{assumption_convex}
that simulating $\ip_\simiter$ will produce
the mode sequence corresponding to the $\region$.
The procedure ends once an
$\ip_\simiter$ is selected that lies outside all
existing $\region$s.
%
%
One drawback of $\CRS$ is that it may
select an input point that is
outside---but still very close to---an
existing $\region$.
Such input points contribute little toward the
total explored volume of $\ipival$.

\subsection{Region Distance Maximization}

We introduce a second
input point selection strategy called \emph{Region Distance Maximization}.
The high-level idea of this strategy is to choose
an input point $\ip_\simiter$
as far away as possible from all existing $\region$s.
We expect that such an $\ip_\simiter$ will either produce
a new mode sequence
(thus aiding in the main goal of maximizing
$|\allms_\nsims|$),
or else contribute significantly toward the volume
of an existing $\region$
(and the total explored volume of $\ipival$).
To choose an input point using $\RDM$,
we require a notion of the distance
from a point to
the existing $\region$s.
In Appendix~\ref{sec_appendix_distance},
we discuss two metrics for calculating the
distance between a candidate input point
$\ip_\simiter$
and a $\region$ $\egregion_\ms$.

\subsection{Rare Mode Sequence Discovery with Accelerated Testing}

Using one of the two input point selection strategies,
we now give the overall algorithm for discovering rare mode sequences.
This procedure is given by
Algorithm \ref{alg_simulate_and_incorporate}.
First, an initial input point from 
$\ipival$
is randomly selected and simulated.
This initial point forms the first $\region$.
Note that on lines
\ref{lst:line:firstregion} and \ref{lst:line:newregion},
the new $\region$ consists of a single point,
so there is no need to take its convex hull.
For each subsequent simulation,
we use either $\CRS$ or $\RDM$
to generate an input point that is both within
$\ipival$
and outside all existing $\region$s.
If simulating the point produces a new mode sequence,
then we create a new $\region$.
Otherwise, we
add the point to the existing $\region$ corresponding to
its mode sequence.
The addition of a new point
expands the $\region$'s convex hull
and contributes toward the
total explored volume of $\ipival$.

\begin{algorithm}[t]
    \caption{Accelerated Testing}
    \begin{algorithmic}[1]
        \label{alg_simulate_and_incorporate}

        \REQUIRE
        CPS simulator
        $\simulate : \mathbb{R}^{\dims} \to \Sigma^{*}$,
        random input point selection function
        \(
            \randip :
            \left[ \mathbb{R}^{\dims}, \mathbb{R}^{\dims} \right]
            \to
            \mathbb{R}^{\dims}
        \),
        input point selector
        $\genip$
        (one of $\CRS$ or $\RDM$),
        simulation budget $\nsims$        
        
        \ENSURE The set $\allms_{\nsims}$ of mode sequences produced by $\nsims$ simulations.

        \STATE \(
            \ip_1 \gets \randip(\ipival),
            \enspace
            \ms \gets \simulate(\ip_1),
            \enspace
            \allms_1 \gets \left\{ \ms \right\}
        \)

        \STATE $\pts_1(\ms) \gets \left\{ \ip_1 \right\}$
        
        \STATE $\regionset_1 \gets \left\{ \pts_1(\ms) \right\}$
        \label{lst:line:firstregion}
        
        \FOR{$\simiter$ = 2 \TO $\nsims$}
        
            \STATE \(
                \ip_\simiter
                \gets
                \genip(\ipival, \regionset_{\simiter-1}),
                \enspace
                \ms \gets \simulate(\ip_\simiter)
            \)
            
            \IF{$\ms \not \in \allms_{\simiter - 1}$}

                \STATE \(
                    \allms_{\simiter}
                    \gets
                    \allms_{\simiter-1} \cup \left\{ \ms \right\}
                \)
            
                \STATE \(
                    \pts_\simiter(\ms)
                    \gets
                    \left\{ \ip_\simiter \right\}
                \)
                
                \STATE \(
                    \regionset_\simiter
                    \gets
                    \regionset_{\simiter-1}
                    \cup
                    \left\{ \pts_\simiter(\ms) \right\}
                \)
                \label{lst:line:newregion}
                
            \ELSE

                \STATE \(
                    \allms_{\simiter}
                    \gets
                    \allms_{\simiter-1}
                \)
            
                \STATE \(
                    \pts_\simiter(\ms)
                    \gets
                    \pts_{\simiter-1}(\ms)
                    \cup
                    \left\{ \ip_{\simiter} \right\}
                \)

                \STATE \(
                    \regionset_{\simiter}
                    \gets
                    \left(
                        \regionset_{\simiter-1}
                        \setminus
                        \left\{ \ch(\pts_{\simiter-1}(\ms)) \right\}
                    \right)
                    \cup
                    \left\{ \ch(\pts_{\simiter}(\ms)) \right\}
                \)
                
            \ENDIF
        \ENDFOR
        
        \RETURN $\allms_{\nsims}$
        
    \end{algorithmic}
\end{algorithm}

\section{Evaluation}

In this section,
we evaluate the use of accelerated testing
for finding rare mode sequences in CPS simulators.
We compare accelerated testing
against the baseline technique
of random sampling
on four CPS benchmarks,
from a synthetic example to a Simulink diagram of a CPS.
To perform the
hull-wise and point-wise distance maximizations
for $\RDM$,
we use the open source tool
ZOOpt\footnote{https://github.com/polixir/ZOOpt}~\cite{liu2017zoopt}.

\begin{figure}[t]
\centering
\includegraphics[width=1.00\columnwidth]{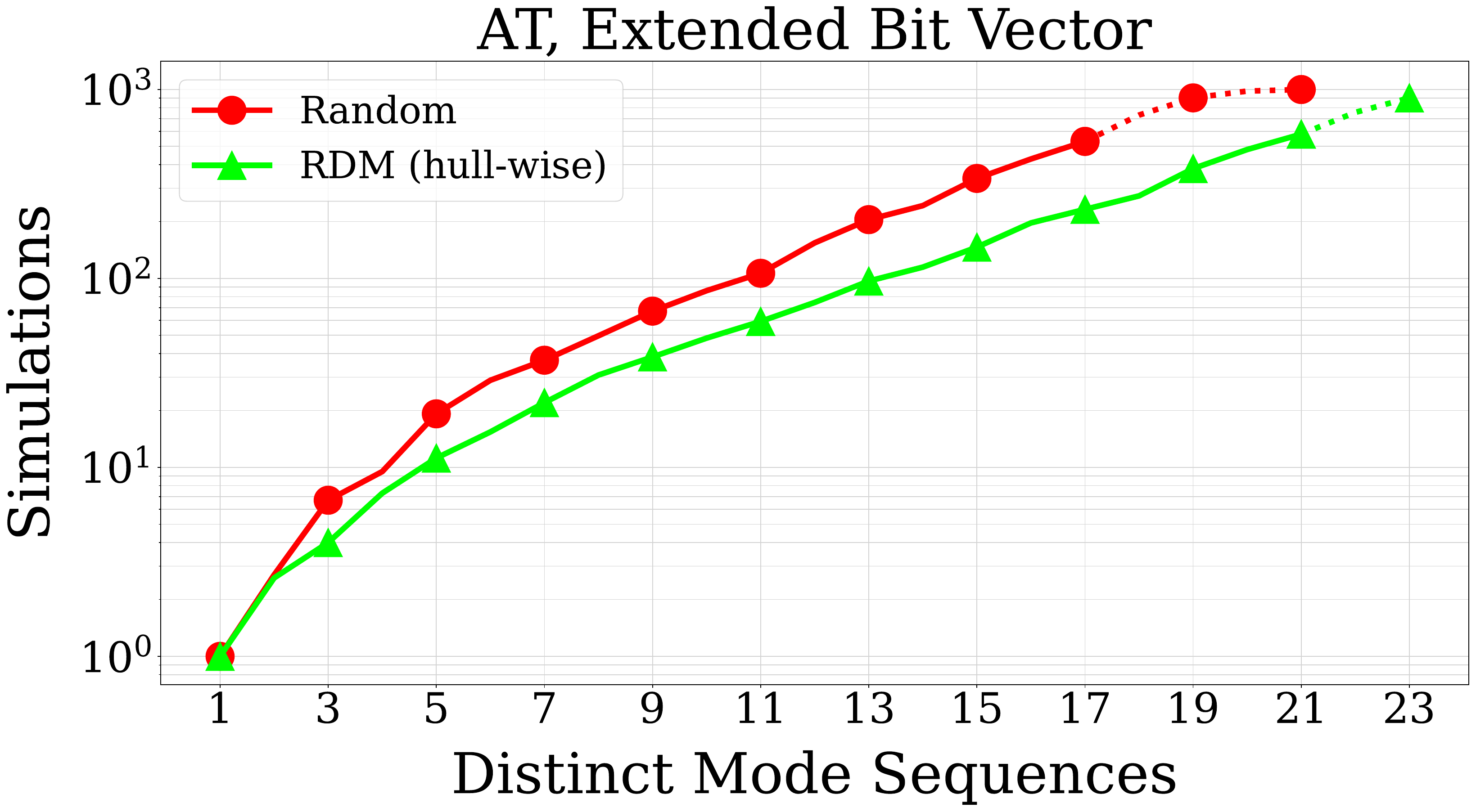}
\caption{
    Results for the $10$-dimensional automatic transmission
    benchmark.
    Using the
    $\atextbitvector$ definition of mode sequence,
    $\RDM$ produced 17 distinct behaviors
    with less than half as many simulations
    as random.
}
\label{fig_results_at}
\end{figure}

Because random sampling and accelerated testing both
select input points non-deterministically,
we take the average over ten trials,
unless noted otherwise.
The curves in
Figures~\ref{fig_results_voronoi},~\ref{fig_results_nav},~\ref{fig_results_gearbox},~and~\ref{fig_results_at}
show the
simulations required to discover
a given number of distinct mode sequences.
Lines become dashed to indicate that at least one
of the ten trials has
already terminated.
Following the trial's termination,
all subsequent mode sequences in that trial
are considered
to be discovered at $\nsims$ simulations,
for the purposes of computing an average.
Note the logarithmic scaling of the simulations axis.



\subsection{Scalable Convex Voronoi Regions}

\begin{figure}
\centering
\includegraphics[width=0.60\columnwidth]{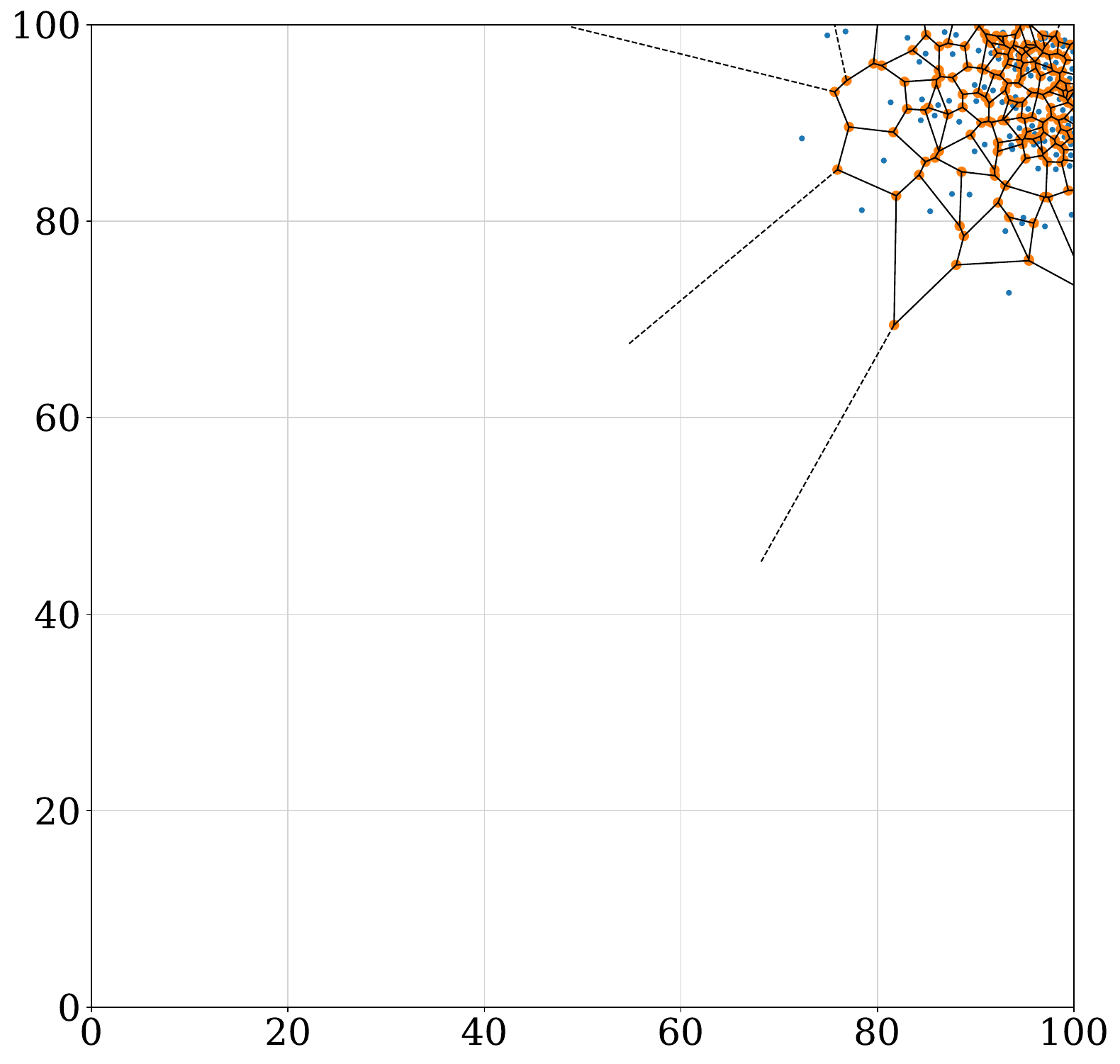}
\caption{
    A 2-dimensional Voronoi diagram created using the given Gaussian distribution,
    representing the input space for the Voronoi system.
    A few large $\region$s take up the majority of the input space,
    while many smaller $\region$s occupy a small portion in the
    top-right.
}
\label{fig_benchmarks_voronoi}
\end{figure}

We first construct a scalable synthetic example system 
to evaluate the improvement in
mode sequence discovery
afforded by our approach.
%
%
The $\dims$-dimensional input space
consists of 100 convex $\region$s,
so $|\allms|=100$.
The size and placement of these $\region$s
are determined by the Voronoi sites.
Since we expect real systems to have mode sequence regions of different sizes, we place the Voronoi sites
using a Gaussian, rather than uniform, distribution.
100 $\dims$-dimensional Voronoi sites $v$ are generated using a truncated normal distribution,
with $v^{(\dimsiter)} \in \left[ 0, 100 \right] \,
\forall \, \dimsiter \in [1, \, \dims], \,
\mu = 100$,
and
$\sigma = 10$.
Figure~\ref{fig_benchmarks_voronoi} shows the approximate sizes and locations of the Voronoi regions under this distribution for 2 dimensions.
To perform a ``simulation,'' we return the unique ID of the Voronoi site closest to the $\dims$-dimensional input. The Voronoi site ID serves as the mode sequence of a simulation.

\begin{figure}[t]
\centering
\includegraphics[width=1.00\columnwidth]{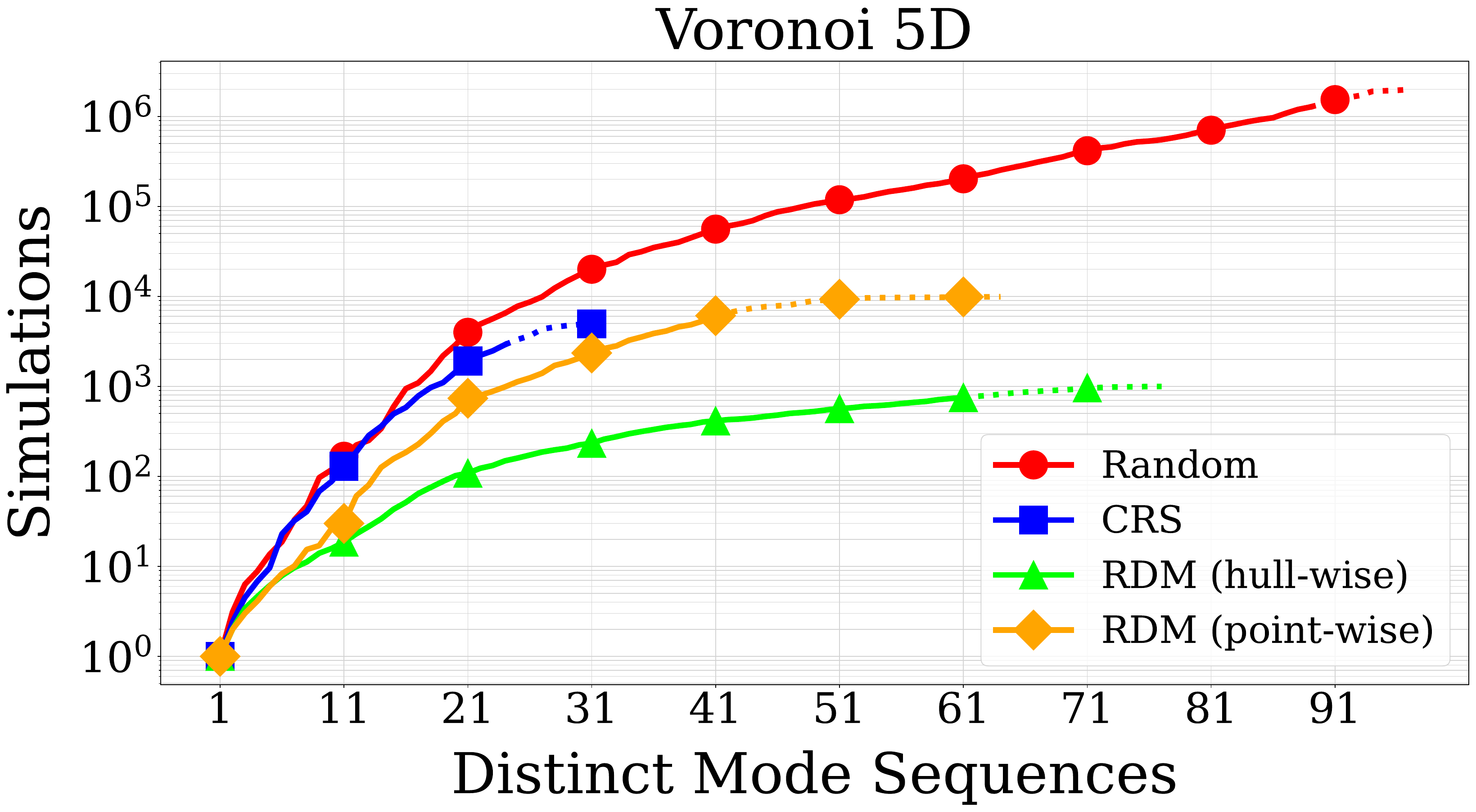}
\caption{
    Results for the Voronoi system.
}
\label{fig_results_voronoi}
\end{figure}

The Voronoi system's input space
has several properties that make
our approach
especially advantageous.
First,
each $\region$ is
inherently convex.
Having convex $\region$s conforms
to Assumption \ref{assumption_convex},
and guarantees that no mode sequences
are missed by skipping input points
inside existing $\region$s.
Second, the specific distribution
from which the Voronoi sites
are selected creates
a low number of large $\region$s,
and many small $\region$s.
Having a few large $\region$s is advantageous
for our approach
because
we are able to skip input points
anywhere inside their large convex hulls.
Compared to uniform random sampling,
our approach spends
more time sampling from the remaining
unexplored portion,
thus discovering more of the small $\region$s.
Figure~\ref{fig_results_voronoi} plots the mode sequences discovered by the four techniques in five dimensions,
where RDM provided an average speedup factor of over 350x.
Additional results for the Voronoi benchmark are given in Appendix~\ref{sec_appendix_vor}.

\subsection{Navigation}

\begin{figure}[t]
\centering
\includegraphics[width=0.6\columnwidth]{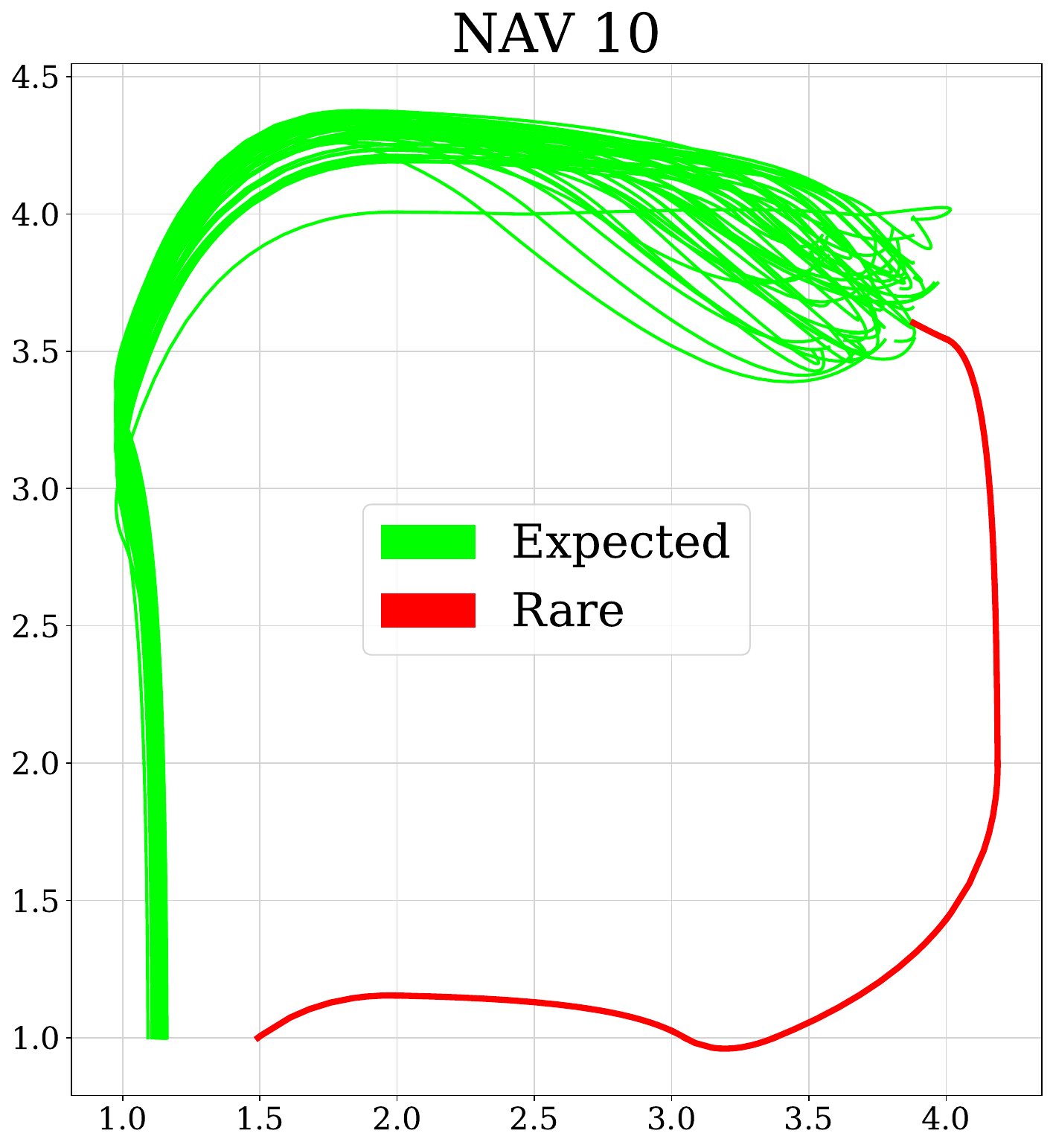}
\caption{
    A rare behavior in one of the navigation benchmarks, where
    the object
    transitions out of the initial cell to the right,
    then proceeds clockwise to the terminal cell.
}
\label{fig_nav10_rare}
\end{figure}

The navigation benchmark (NAV)
models an object’s movement through the plane~\cite{fehnker2004}.
The plane is divided into a grid of $1 \times 1$ cells with unique IDs,
where each cell is encoded with one of eight possible desired velocities.
While inside a cell, the object experiences linear dynamics
that bring its velocity toward that cell's desired velocity.
The 4-dimensional continuous state consists of the object's velocity and position in the plane.
The object begins within a given interval $\ipival$ of
4-dimensional initial states in the grid.
The discrete state of the system is the ID of the current cell.
Discrete state transitions occur when the object touches the boundary of a neighboring grid cell.
Some cells are designated as \textit{terminal cells}; simulation ends once the object transitions to a terminal cell.
The mode sequence of a simulation is the ordered list of cells that the object visits.
A cell may appear in the mode sequence more than once.
%
%
%
Figure~\ref{fig_nav10_rare} shows one of the rare
behaviors in the NAV 10 benchmark, where
the object exits the initial mode to the right
instead of the left.

\begin{figure}[t]
\centering
\includegraphics[width=1.00\columnwidth]{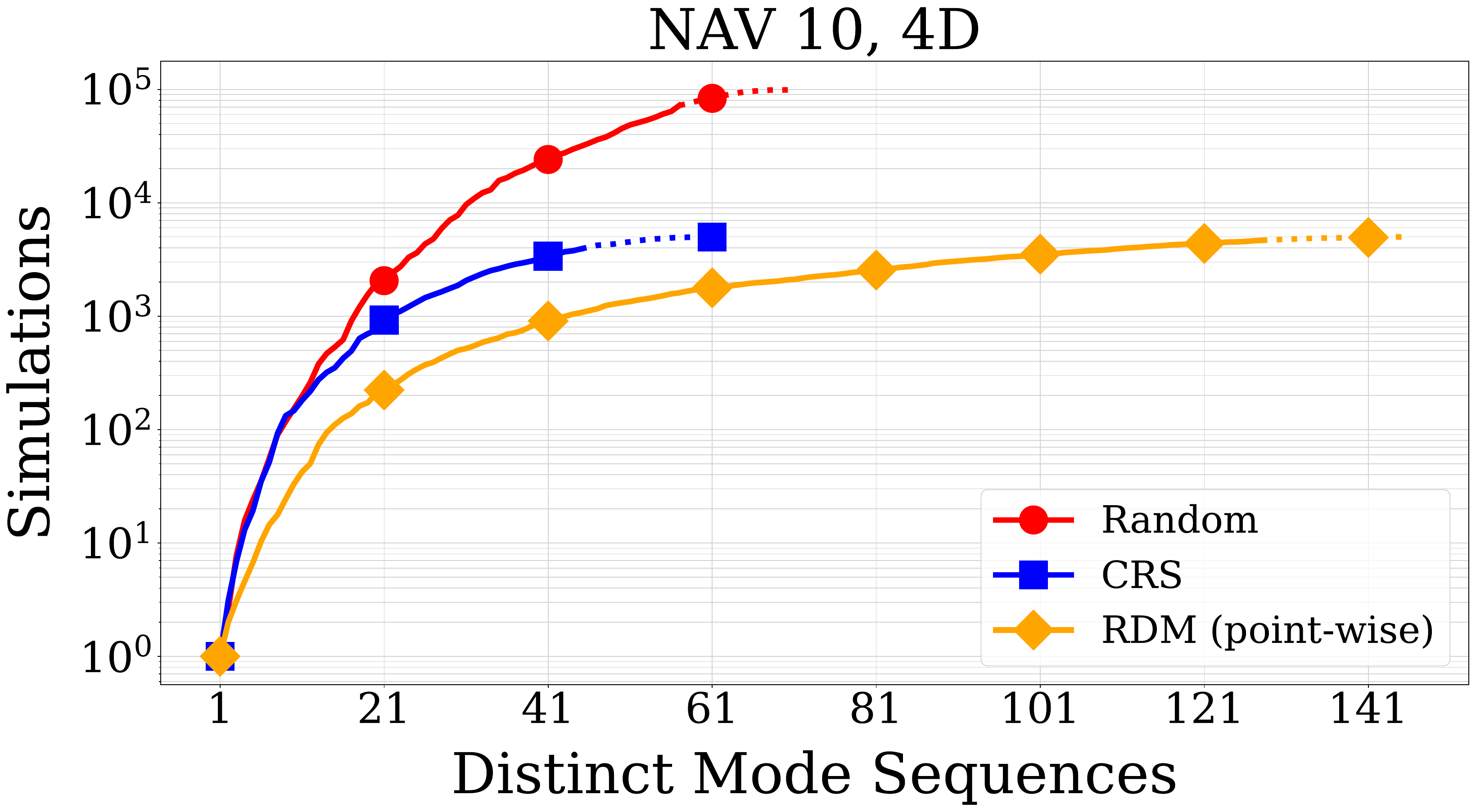}
\caption{
Mode sequence discovery rates
for the NAV 10 benchmark.
$\RDM$ was able to discover
more than twice as many distinct behaviors
in 20x fewer simulations
than random sampling.
}
\label{fig_results_nav}
\end{figure}

Figure~\ref{fig_results_nav} shows
mode sequence discovery in
the NAV 10 benchmark.
Accelerated testing discovered an average of
over 136 mode sequences in 5000 simulations, while
random testing failed to discover even half as
many mode sequences in 100000 simulations.
%
%
Additional results for other NAV benchmarks appear
in Appendix~\ref{sec_appendix_nav}.
Across the
16 NAV benchmarks analyzed,
$\CRS$ provided an average speedup of 2.63x,
and
$\RDM$ provided an average speedup of 6.24x.


\subsection{Gearbox Meshing}

\begin{figure}[t]
\centering
\includegraphics[width=1.00\columnwidth]{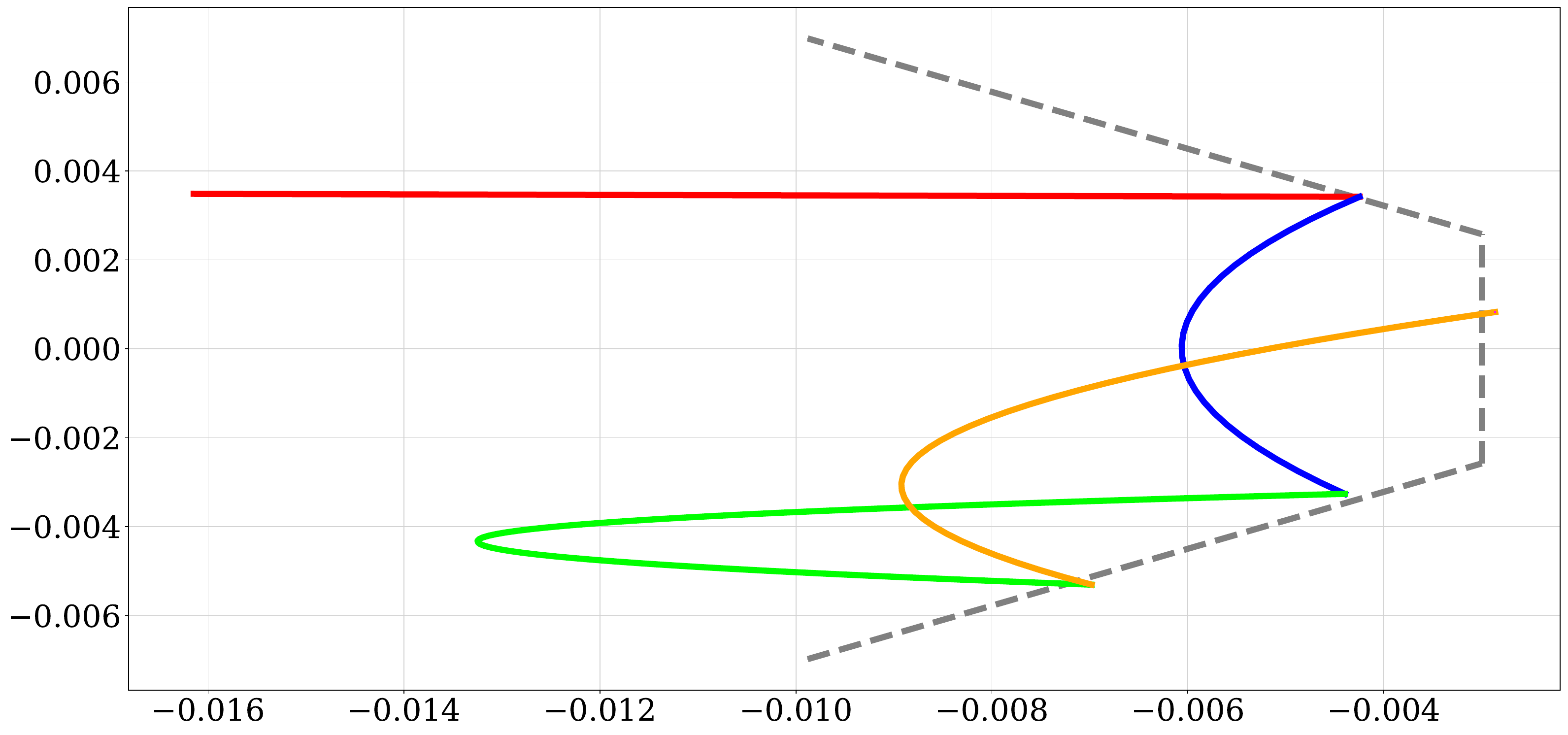}
\caption{
    An example trajectory of the gearbox meshing benchmark.
    The sleeve bounces once against the upper tooth,
    then twice against the lower tooth,
    before successfully meshing with the gear;
    thus, the mode sequence of this simulation is
    $\texttt{2,1,1,3}$.
}
\label{fig_diagram_gearbox}
\end{figure}

\begin{figure}[t]
\centering
\includegraphics[width=1.00\columnwidth]{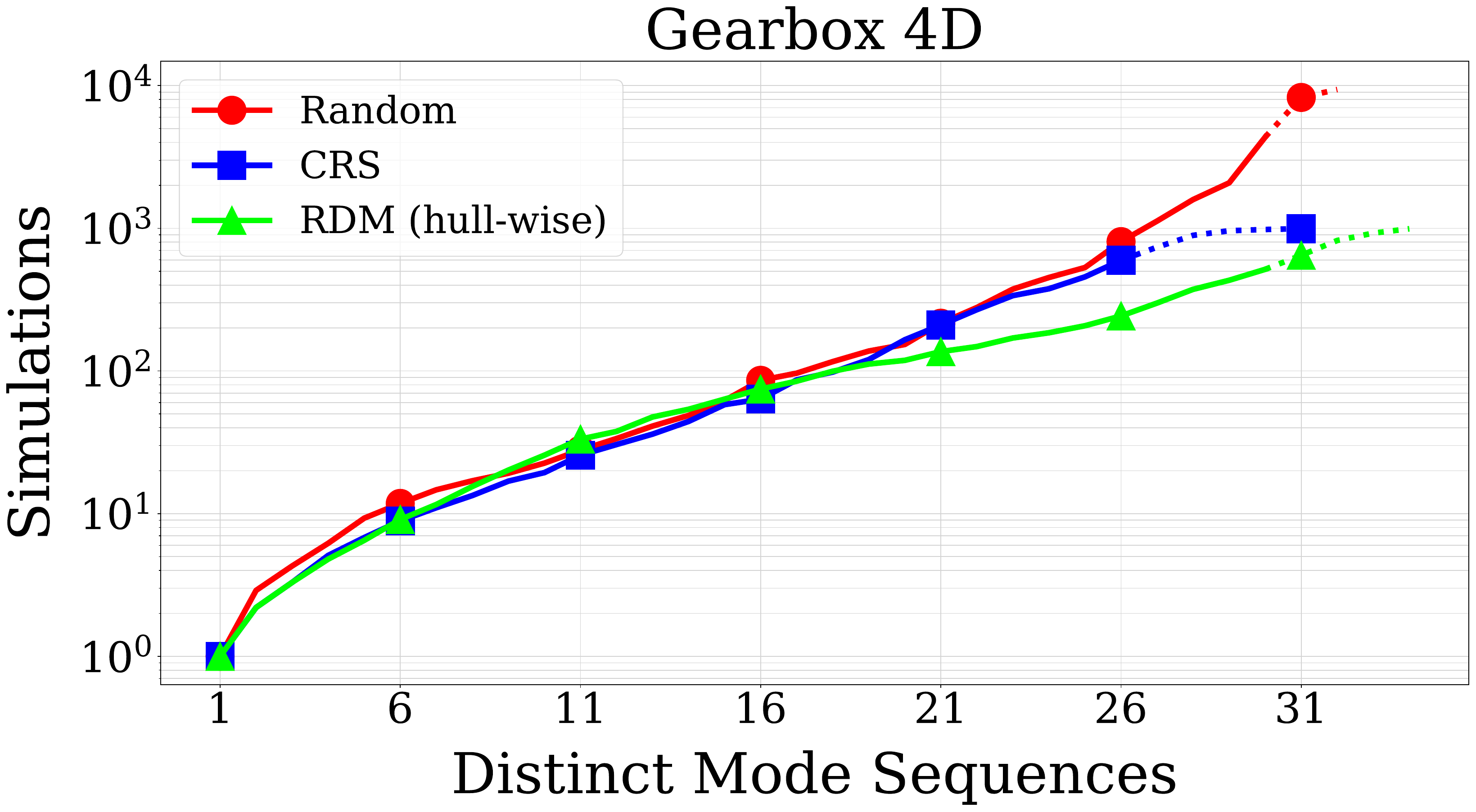}
\caption{
    Results for the 4D
    gearbox meshing benchmark.
    $\RDM$
    testing on average found novel mode sequences
    that random simulations never produced.
}
\label{fig_results_gearbox}
\end{figure}

The gearbox meshing benchmark (Gearbox) models the meshing of a sleeve with a gear during automotive motor-transmission from first to second gear~\cite{chen2014motor}. If the sleeve is properly aligned with the gear when they meet, then the meshing process will complete successfully. Otherwise, the sleeve will collide with a gear tooth, bounce away, and reattempt the meshing process.
%
The sleeve is modeled by a point in the 4-dimensional plane, where
the position and velocity of the sleeve relative to the gear comprise the continuous state.
The discrete state is one of \texttt{meshed} or \texttt{free}.
%
%
Discrete transitions occur whenever the sleeve collides with the upper or lower gear
tooth---altering the sleeve’s velocity and
accumulating impact impulse---as well as
once the meshing process completes.
The simulation ends once the \texttt{meshed} state is reached.
%
The sequence of discrete transitions
serves as the mode sequence.
We refer to the transition
$\texttt{free} \to \texttt{free}$ as \texttt{1} when the
sleeve bounces against the lower tooth, and \texttt{2} when
the sleeve bounces against the upper tooth.
We refer to the transition $\texttt{free} \to \texttt{meshed}$
as \texttt{3}.
%
%
An example simulation and its corresponding mode sequence are
shown in Figure \ref{fig_diagram_gearbox}.
%
Compared to the original benchmark,
we choose the expanded input interval
$p_x \in [-0.017, \, -0.016]$,
$p_y \in [-0.005, 0.005]$
proposed in~\cite{duggirala2019emsoft}
to allow for a larger number of distinct behaviors.
Additionally, we allow the initial velocities, $v_x$ and $v_y$,
to take on a range of values,
$v_x \in [-0.2364, 0.2364]$,
$v_y \in [-0.1260, 0.1260]$.
%
%
Figure \ref{fig_results_gearbox} plots the number of mode sequences
discovered by each of the input point selection strategies.
$\RDM$ provided more than an 8x improvement, finding
30 mode sequences in 515.8 simulations
compared to 4358.4 random simulations.

\subsection{Automatic Transmission}

The Automatic Transmission benchmark (AT) simulates a vehicle equipped with an automatic transmission system~\cite{hoxha2014benchmarks}.
The system's continuous states are the engine speed $\omega$ (RPM) and vehicle velocity $v$ (mph). 
The discrete state of the system is the gear $g_\alpha$ for $\alpha \in [\texttt{1}, \, \texttt{4}]$.
The system is $\dims$-dimensional and depends deterministically on two control inputs, throttle and brake, which are piecewise constant over $(\dims/2)$ fixed intervals.
%
%
%
In the experiments below,
we select $5$ pairs of throttle and brake inputs, applied every 6 seconds throughout the 30-second simulation. This amounts to an $\dims$-dimensional input space for the system,
with $\dims=10$.

We define the mode sequence
for the AT benchmark using an $\atextbitvector$
of 13 pass/fail bits.
We consider 13 STL safety specifications
based on the STL properties from~\cite{hoxha2014benchmarks}.
The mode sequence of a simulation
is the 13-bit vector
consisting of \texttt{1}s and \texttt{0}s,
determined by
whether each safety specification was satisfied.
Although this mode sequence definition uses information besides
discrete modes,
it only requires simulation outputs
that would be available from a black-box simulator.
The 13 STL specifications are:
\begin{equation*}
    \begin{tabular}{ll}
        \(
            \square_{[0, \, 10]} \,
            v < \Bar{v},
        \quad
            \Bar{v} \in \{ 80, 85, 90, 95 \}
        \) & (four bits)
        \\
        \(
            \square_{[0, \, 8]} \,
            \omega < \Bar{\omega},
        \quad
            \Bar{\omega} \in \{ 4500, 4600, 4700 \}
        \) & (three bits)
        \\
        \(
            \square_{[0, \, 30]} \,
            \omega < 2000
            \rightarrow
            \square_{[0, \, 8]} \,
            v < \Bar{v},
        \) & \\
        \qquad \qquad \(
            \Bar{v} \in \{ 80, 100 \}
        \) & (two bits)
         \\
        \(
            \square_{[0, \, 30]} \,
            \big(
            ( \lnot g_\gearnum \wedge \circ \, g_\gearnum )
            \rightarrow
            \circ \, \square_{[0, 1]} \, g_\gearnum
            \big),
        \) \\
        \qquad \qquad \(
            \gearnum \in \{ 1, 2, 3, 4 \}
        \) & (four bits)
    \end{tabular}
\end{equation*}
where $\circ \, \varphi \equiv \diamond_{[0.001,\, 0.1]} \, \varphi$.
For the
$\atextbitvector$ definition of
mode sequence,
$\RDM$ discovered
an average of 22 mode sequences,
whereas random sampling averaged
18.2 in the same number of simulations.
All ten trials produced
at least 17 mode sequences;
however,
accelerated testing with
$\RDM$ required only
231.9 simulations, compared to 530.7 simulations
with random sampling.

\section{Discussion and Future Work}

The results indicate that considering convex
mode sequence regions
constructed from simulated input points
is a promising strategy for rapidly discovering
new mode sequences
(and thus, unprecedented behaviors)
in CPS simulators.
Many system characteristics can affect the
performance of our approach.
%
%
%
%
Depending on the inherent convexity
of mode sequence $\region$s in the input space,
accelerated testing could sometimes
miss
interesting areas with many distinct mode sequences.
%
%
We therefore expect accelerated testing to perform well
on systems with inherently convex mode sequence
$\region$s,
such as the Voronoi benchmark.
In the future,
we intend to devise a metric that
estimates the inherent convexity of a hybrid
system's $\region$s before simulating.
%
Occasionally sampling within existing regions could
also act as a safeguard
against non-convex regions.

Another factor that influences the performance of our approach
is the relative sizes of the $\region$s.
Accelerated testing has the greatest impact when
there are a few large $\region$s,
and many smaller regions.
The early input points quickly approach the hulls
of the large $\region$s,
preventing many redundant simulations.
On the other hand,
when all $\region$s are of uniform size,
random sampling will
quickly uncover all the distinct
mode sequences.
Another metric we intend to develop will
estimate a system's distribution of
$\region$ sizes and predict the improvement
afforded by accelerated testing.

The dimension of a system also contributes
to our approach's mode sequence discovery rate.
In higher dimensions,
the volume of the input space
becomes exponentially larger.
Furthermore, many more input points
are needed in order to enclose
the convex hull of a $\region$.
$\CRS$ provided almost
no improvement in 10 dimensions or higher,
since very few of the first several thousand
input points were inside an existing $\region$.
The $\RDM$ approach quickly
explores portions of the input space far from
previous input points.
It is therefore less reliant on existing
$\region$s with large volume, and tends
to discover mode sequences than
$\CRS$, particularly in higher dimensions.

\section{Conclusion}
In this work, we proposed a new goal for simulation-based analysis of cyber-physical systems: finding rare behaviors by analyzing mode sequence outputs of a black-box CPS simulator.
We hypothesize that rare behaviors often correspond with unknown unknowns---unanticipated problems that can manifest in a complex system.
As an engineer's time is limited and expensive, our method identifies the most interesting situations for manual review.
We proposed two algorithms, Convex Rejection Sampling and Region Distance Maximization, that accelerate the process of finding these rare behaviors, in some cases by over two orders of magnitude.
As our approach does not require a specification be provided, we believe it can be a complementary tool in a comprehensive CPS testing framework, which includes other approaches like manual feature testing, regression testing, and falsification methods.

\begin{acks}
This material is based upon work supported by the Office of Naval Research under award number N00014-22-1-2156, and the National Science Foundation under Award No. 2237229. Any opinions, findings, and conclusions or recommendations expressed in this material are those of the author(s) and do not necessarily reflect the views of the United States Navy.
\end{acks}

\bibliographystyle{IEEEtran}
\bibliography{bibs/bak,bibs/bibliography}

\newpage

\appendix
\section{Point-to-Region Distance Metrics}
\label{sec_appendix_distance}

The first metric,
which we call the
\textit{hull-wise distance},
measures the distance from $\ip_\simiter$
to the convex hull of a $\region$.
In order to formulate the hull-wise distance calculation
as an $\linprog$,
we seek the smallest radius $\cuberad \in \mathbb{R}$
such that an axis-aligned hypercube,
with radius $\cuberad$ and
center $\ip_\simiter$,
intersects the $\region$.
Note that the intersection can be
any convex combination of the
previously simulated input points.
We represent the axis-aligned
hypercube $\hypcube$
centered at $\cubecenter \in \mathbb{R}^{\dims}$
with radius $\cuberad \in \mathbb{R}$
as
\begin{equation}
    \hypcube(\cubecenter, \, \cuberad)
    =
    \{ \cubept \in \mathbb{R}^{\dims}
    \mid
    \cubecenter - \cuberad \, \unitvec
    \leq
    \cubept
    \leq
    \cubecenter + \cuberad \, \unitvec
    \}
\end{equation}
where $\unitvec$
is the $\dims$-dimensional vector of all ones;
 $\cuberad \, \unitvec$ represents
scalar-vector multiplication;
and $a \leq b$
performs element-wise comparison,
satisfied only if
$a^{(\dimsiter)} \leq b^{(\dimsiter)} \enspace \forall \dimsiter \in [1, \dims]$.
We calculate the hull-wise distance between an
input point $\ip_\simiter$ and
a $\region$ $\egregion_\ms$:
\begin{equation}
    \label{eq_hull_dist}
    \begin{tabular}{ll}
            $\hulldist(\ip_\simiter, \, \egregion_\ms)=$
            &
            $\min \cuberad$
            \\
            $\st$
            &
            \(
                \hypcube(\ip_\simiter, \, \cuberad)
                \cap
                \egregion_\ms
                \neq \emptyset
            \)
    \end{tabular}
\end{equation}

The second distance metric is
\textit{point-wise distance}.
The point-wise distance between $\ip_\simiter$
and $\region$ $\egregion_\ms$ is
the squared distance between $\ip_\simiter$
and the nearest previously simulated input point
$\nearestpt\in\pts_{\simiter-1}(\ms)\subseteq\egregion_\ms$.
Assume that $\region$ $\egregion_\ms$
has access to the data structure containing the
mapping from $\ms$ to $\pts_{\simiter-1}(\ms)$.
\begin{equation}
    \label{eq_point_dist}
    \ptdist(\ip_\simiter, \, \egregion_\ms)
    =
    \min\limits_{\nearestpt \in \pts_{\simiter-1}(y)}
    \sum_{\dimsiter=1}^{\dims}
    \left(
    \ip_{\simiter}^{(\dimsiter)}
    -
    \nearestpt^{(\dimsiter)}
    \right)^2
\end{equation}

We wish to optimize over the input space $\ipival$
in search of an input point $\ip_\simiter$ whose
distance from the closest $\region$ is as far
as possible.
%
%
Formally, we seek the $\ip_\simiter$
returned by the following optimization:
\begin{equation}
    \label{eq_cost}
    \objfunc(\ipival, \, \regionset_{\simiter-1}) =
    \argmax_{\ip_\simiter \in \ipival}
    \min_{\egregion_\ms \in \regionset_{\simiter-1}}
    \dist(\ip_\simiter, \, \egregion_\ms)
\end{equation}
where $\dist$ is one of $\hulldist$ or $\ptdist$.
To solve the maximization problem,
we use zeroth-order optimization,
also known as derivative-free
or black-box optimization.
Given a budget of $\optbudget$ optimization iterations,
the solution is the input point $\ip_\simiter$
with the largest distance to its closest $\region$.
We perform a convex hull inclusion check (an $\linprog$)
per $\region$ for each intermediate solution,
to ensure that the
input point lies outside all existing $\region$s.
%

A discussion of the trade-offs between the two distance metrics follows.
One disadvantage of performing optimization using
the hull-wise distance metric is that for every
input point $\ip_\simiter$,
$\RDM$ must perform $\optbudget$ $\linprog$s
(one at each optimization iteration, to calculate the hypercube radius
of each intermediate solution).
%
%
However, hull-wise distance has the benefit that
with a large enough optimization budget $\optbudget$,
the returned solution approximates the global optimum.
%

The point-wise distance metric offers the advantage of
faster computation time by avoiding the need to
solve an $\linprog$ for each intermediate solution.
%
%
Instead, the primary calculation required to find the
point-wise squared distance is a simple dot product
$\distmat \cdot \distmat^{\transpose}$, where
$\distmat$ represents the difference between
the intermediate solution $\ip_\simiter$ and
each of the previously simulated input points:
\begin{equation}
    \distmat =
    \begin{bmatrix}
        \ip_1^{(1)} - \ip_{\simiter}^{(1)} &
        \dots &
        \ip_{\simiter - 1}^{(1)} - \ip_{\simiter}^{(1)}
        \\
        \vdots &
        \ddots &
        \vdots
        \\
        \ip_1^{(\dims)} - \ip_{\simiter}^{(\dims)} &
        \dots &
        \ip_{\simiter - 1}^{(\dims)} - \ip_{\simiter}^{(\dims)}
    \end{bmatrix}
\end{equation}
The disadvantage of performing optimization
with point-wise distance
is that it might converge to
an $\ip_\simiter$ that
is inside an existing $\region$.
Take for example a $\region$ that is a large
$\dims$-dimensional simplex.
Using point-wise distance, $\RDM$ might return the center of this
simplex as its optimal solution, because it is technically far from
all previously simulated input points.
To prevent this solution from being returned in future optimization attempts,
when using point-wise distance,
we insert every input point returned by the optimizer
into the $\region$, but we only simulate input points
when they lie outside all existing regions.

\section{Additional Evaluation}

Tables~\ref{tab_vor}~and~\ref{tab_nav_onlytenmsormore}
give the number of distinct mode sequences
$|\allms_\nsims|$ found by accelerated testing
within a simulation
budget of $\nsims$.
We compare this against the average number of
random simulations required to find
this number of mode sequences.
In cases where accelerated testing averages to a non-integer number
of mode sequences,
we consider the number of random simulations required to find the floor,
$\big\lfloor |\allms_{\nsims}| \big\rfloor$.
In these cases,
the listed speedup factor is a conservative estimate,
since the extra mode sequence discovered by some accelerated
testing trials could be exceedingly rare.

The speedup factor is
calculated as:
\begin{equation}
    \text{Speedup Factor}
    =
    \frac{\text{Avg. Rand Sims}}{\nsims}
\end{equation}
Note that we are underapproximating the true speedup,
as accelerated testing may have
discovered the $|\allms_\nsims|^{\text{th}}$ mode sequence
earlier than the $\nsims^{\text{th}}$ simulation,
whereas random simulations were halted
immediately upon finding the
$|\allms_\nsims|^{\text{th}}$ mode sequence.

\subsection{Scalable Convex Voronoi Regions}
\label{sec_appendix_vor}

\begin{table}[t]
    \centering
    \caption{Voronoi Mode Sequence Discovery}
    \label{tab_vor}
    \begin{tabular}{@{}llllll@{}}
  \hline
    & \multicolumn{1}{c}{$\nsims$}
    & \multicolumn{1}{l}{$|\allms_{\nsims}|$}
    & \multicolumn{1}{l}{Random}
    & \multicolumn{1}{l}{Speedup} \\
\hline

${\dims=2}$ \\
\quad $\CRS$ & 1000 & 95.2 & 16329.5 & 16.3 \\
 \quad $\RDM$ (hull) & 300 & 90.6 & 10830.8 & 36.1 \\
 \quad $\RDM$ (point) & 1200 & 99.7 & 30551.2 & 25.5 \\
    
   ${\dims=3}$ \\ 
\quad$\CRS$ & 2000 & 75.1 & 22133.4 & 11.1 \\
\quad $\RDM$ (hull) & 500 & 88.3 &  46719.5 & \textbf{{\color{red}93.4}} \\
\quad $\RDM$ (point) & 2500 & 95 &  84934.2 & 34.0 \\
    
    ${\dims=5}$ \\
\quad $\CRS$ & 5000 & 27.8 &  9907.2 & 1.98 \\
\quad$\RDM$ (hull) & 1000 & 69.8 & 354345.9 & \textbf{{\color{red}354.3}} \\
\quad$\RDM$ (point) & 10000 & 50.5 & 111991.1 & 11.2 \\
    
 ${\dims=10}$ \\   
\quad$\CRS$ & 1000 & 24.4 &  765.3 & 0.77 \\
\quad $\RDM$ (hull) & 1000 & 51 &  369151.7 & \textbf{{\color{red}369.2}} \\
\quad $\RDM$ (point) & 10000 & 40.1  & 56578.3 & 5.7 \\
    \hline
    \end{tabular}
\end{table}

\begin{figure}[t]
\centering
\includegraphics[width=1.00\columnwidth]{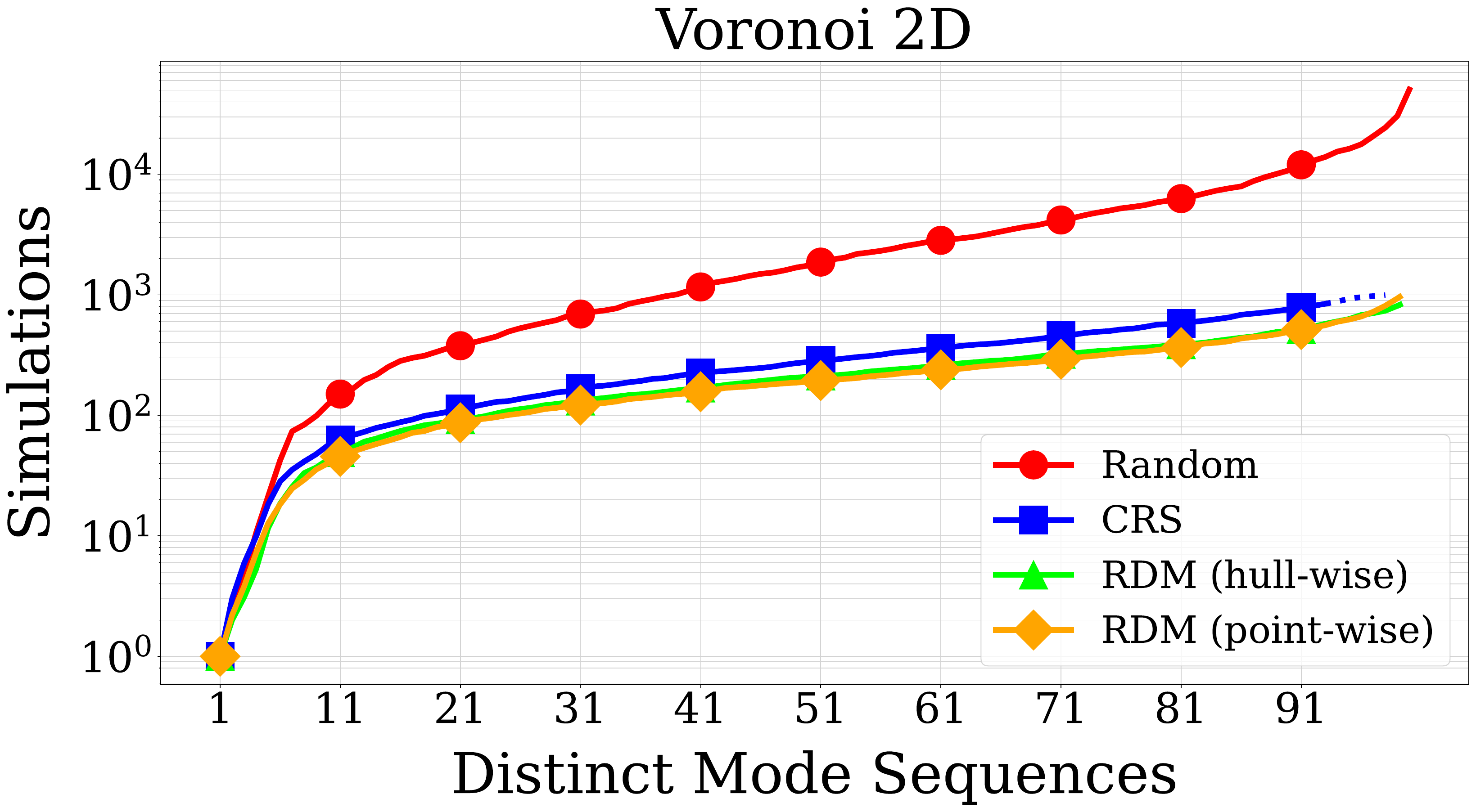}
\includegraphics[width=1.00\columnwidth]{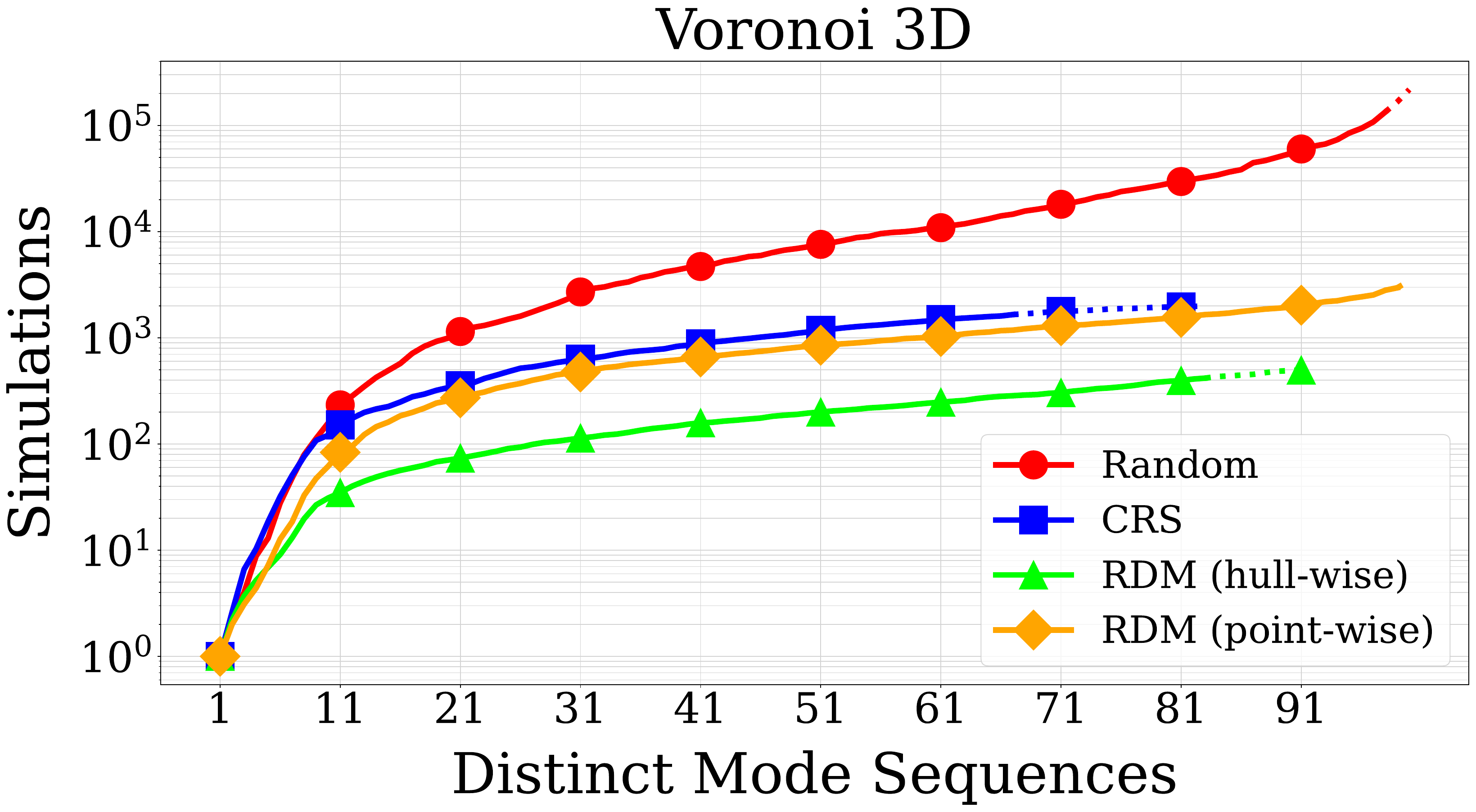}
\includegraphics[width=1.00\columnwidth]{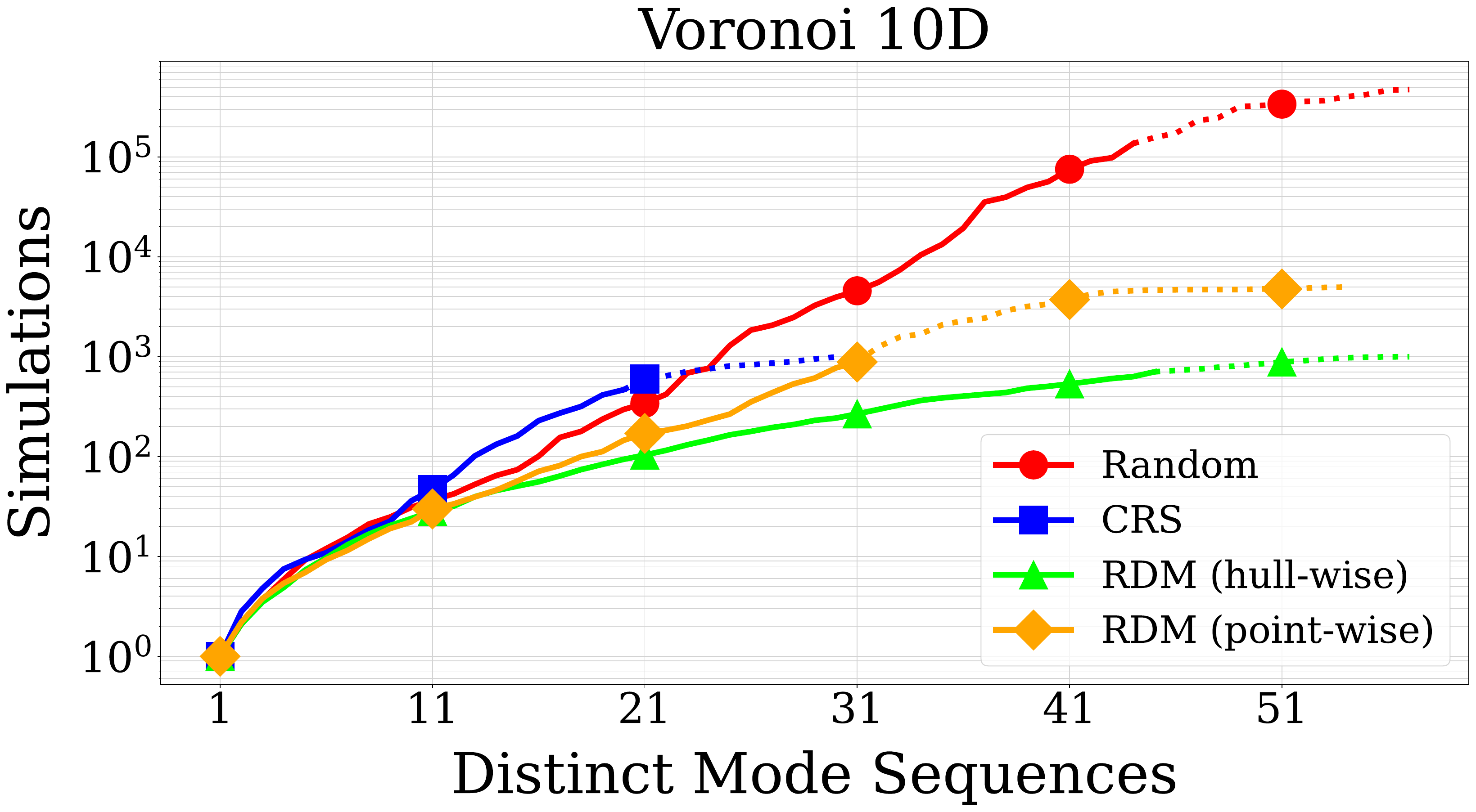}
\caption{
    Additional results for the Voronoi system.
}
\label{fig_results_voronoi_2_3_10}
\end{figure}

Table~\ref{tab_vor} lists the full results for each input point selection strategy in the synthetic scalable Voronoi regions benchmark.
Figure~\ref{fig_results_voronoi_2_3_10} plots the
average number of mode sequences discovered.
All four strategies struggled when
searching for mode sequences in 10 dimensions, finding roughly half as many mode sequences (about 50)
as in lower dimensions (about 90).
However, the RDM method was still over 300 times faster in this case compared with 
random sampling.
The reduced performance of all methods can be explained by the
curse of dimensionality.
For example, in 10 dimensions,
we would need
$2^{10} = 1024$ input points
to enclose the convex hull of
a hypercube,
whereas in 2 dimensions,
we would only need
$2^2 = 4$ input points.
The number of rejections experienced during
$\CRS$ per successful simulation is nearly
zero in 10 dimensions.

\subsection{Navigation}
\label{sec_appendix_nav}

Table~\ref{tab_nav_onlytenmsormore} provides
details about the exact
speedup factor for each NAV benchmark.
Note that all $\RDM$ trials
for the NAV benchmarks
use the point-wise distance metric.
In the evaluations,
we omit NAV benchmarks for which we witnessed
ten or fewer distinct mode sequences,
since random testing tends to be sufficient
for finding the behaviors
in a low number of simulations.

\begin{table}[t]
    \caption{Navigation Mode Sequence Discovery}
    \label{tab_nav_onlytenmsormore}
    \centering
{\renewcommand{\arraystretch}{1.0}
    \begin{tabular}{@{}lllll@{}}
\hline
    & \multicolumn{1}{c}{$\nsims$}
    & \multicolumn{1}{l}{$|\allms_{\nsims}|$}
    & \multicolumn{1}{l}{Rand Sims}
    & \multicolumn{1}{l}{Speedup} \\
\hline

\textbf{NAV 10} & & & & \\
\quad $\CRS$	 & 	5000	 & 	54.5	 & 	56519.5	 & 	11.30 \\
\quad $\RDM$	 & 	2000	 & 	57 \footnotemark	 & 	72133.6	 & 	\textbf{{\color{red}36.07}} \\ 

\textbf{NAV 11} & & & & \\
\quad $\CRS$	 & 	5000	 & 	63	 & 	23494.9	 & 	4.70 \\ 
\quad $\RDM$	 & 	5000	 & 	84.2	 & 	72779.3	 & 	\textbf{{\color{red}14.56}} \\

\textbf{NAV 16} & & & & \\
\quad $\CRS$	 & 	5000	 & 	88.8	 & 	7445.2	 & 	1.49 \\
\quad $\RDM$	 & 	5000	 & 	95.7	 & 	19634.2	 & 	3.93 \\ 
\textbf{NAV 17} & & & & \\
\quad $\CRS$	 & 	5000	 & 	60.1	 & 	11837.1	 & 	2.37 \\ 
\quad $\RDM$	 & 	5000	 & 	63.2	 & 	18061.1	 & 	3.61 \\ 
\textbf{NAV 18} & & & & \\
\quad $\CRS$	 & 	5000	 & 	130.1	 & 	10262.9	 & 	2.05 \\ 
\quad $\RDM$	 & 	5000	 & 	134.6	 & 	14113.5	 & 	2.82 \\

\textbf{NAV 20} & & & & \\
\quad $\CRS$	 & 	2500	 & 	38.2	 & 	7093.4	 & 	2.84	 \\ 
\quad $\RDM$	 & 	2000	 & 	38.2	 & 	7093.4	 & 	3.55	 \\ 
\textbf{NAV 21} & & & & \\
\quad $\CRS$	 & 	5000	 & 	83.8	 & 	14249	 & 	2.85	 \\ 
\quad $\RDM$	 & 	5000	 & 	89	 & 	20258.3	 & 	4.05	 \\ 
\textbf{NAV 22} & & & & \\
\quad $\CRS$	 & 	5000	 & 	147.3	 & 	15678.6	 & 	3.14	 \\ 
\quad $\RDM$	 & 	5000	 & 	192.6	 & 	49203.1	 & 	\textbf{{\color{red}9.84}}	 \\ 

\textbf{NAV 23} & & & & \\
\quad $\CRS$	 & 	5000	 & 	399.6	 & 	7655.7	 & 	1.53	 \\ 
\quad $\RDM$	 & 	3500	 & 	386.7	 & 	7120.8	 & 	2.03	 \\ 
\textbf{NAV 24} & & & & \\
\quad $\CRS$	 & 	5000	 & 	857.1	 & 	7618.3	 & 	1.52	 \\ 
\quad $\RDM$	 & 	5000	 & 	901.2	 & 	8287.4	 & 	1.66	 \\ 
\textbf{NAV 25} & & & & \\
\quad $\CRS$	 & 	1000	 & 	31.4	 & 	1844.3	 & 	1.84	 \\ 
\quad $\RDM$	 & 	5000	 & 	48	 & 	23869.6	 & 	4.77	 \\ 
\textbf{NAV 26} & & & & \\
\quad $\CRS$	 & 	1000	 & 	54.4	 & 	1352.8	 & 	1.35	 \\ 
\quad $\RDM$	 & 	5000	 & 	92.1	 & 	18085.2	 & 	3.62	 \\ 
\textbf{NAV 27} & & & & \\
\quad $\CRS$	 & 	1000	 & 	109.8	 & 	1168.5	 & 	1.17	 \\ 
\quad $\RDM$	 & 	5000	 & 	208.9	 & 	11778.8	 & 	2.36	 \\ 
\textbf{NAV 28} & & & & \\
\quad $\CRS$	 & 	5000	 & 	116.5	 & 	8945	 & 	1.79	 \\ 
\quad $\RDM$	 & 	5000	 & 	126.3	 & 	17894.1	 & 	3.58	 \\ 
\textbf{NAV 29} & & & & \\
\quad $\CRS$	 & 	1000	 & 	188.9	 & 	1020	 & 	1.02	 \\ 
\quad $\RDM$	 & 	10000	 & 	535.8	 & 	20295.6	 & 	2.03	 \\ 
\textbf{NAV 30} & & & & \\
\quad $\CRS$	 & 	7500	 & 	790.9	 & 	8811.3	 & 	1.17	 \\ 
\quad $\RDM$	 & 	5000	 & 	733.6	 & 	7037.6	 & 	1.41	 \\ 

\hline
    \end{tabular}
}
\end{table}
\footnotetext{
Due to the amount of computation time
required for random testing,
we only consider the first
57 mode sequences discovered by
$\RDM$ in NAV 10.
To find the full 136 mode sequences that
$\RDM$ discovered, a single random
trial required multiple days and more than
800000 simulations.
We were therefore unable to
compute the precise speedup factor for the full 136 mode sequences,
but we predict that it would be around 160 times.
}

\end{document}